\documentclass[conference]{IEEEtran}
\IEEEoverridecommandlockouts
\usepackage{cite}
\ifCLASSINFOpdf
  \usepackage[pdftex]{graphicx}
  \DeclareGraphicsExtensions{.pdf,.jpeg,.png}
\else
  \usepackage[dvips]{graphicx}
  \DeclareGraphicsExtensions{.eps}
\fi
\usepackage[cmex10]{amsmath}
\usepackage{amssymb}  
\usepackage{amsmath}  
\usepackage{steinmetz}
\graphicspath{{./figures/}}
\usepackage{amsthm}

\usepackage{algorithm,algorithmic}

\begin{document}

\title{Cram\'{e}r-Rao Bounds for the Simultaneous Estimation of Power System Electromechanical Modes and Forced Oscillations \\
\thanks{This material is based upon work supported by the National Science Foundation under Grant No. 1944689}
}

\author{\IEEEauthorblockN{Luke Dosiek}
\IEEEauthorblockA{Department of Electrical, Computer \& Biomedical Engineering\\
Union College\\
Schenectady, NY 12309\\
Emails: dosiekl@union.edu}
}

\maketitle

\begin{abstract}
In this paper, the Cram\'{e}r-Rao Bounds (CRB) for the simultaneous estimation of power system electromechanical modes and forced oscillations (FO) are derived. Two cases are considered; in the first case only the steady-state response to the FO is present in the measured system output used by estimation algorithms. In the second, the startup transient of the FO is present in addition to the steady-state response. The CRBs are analyzed numerically to explore sensitivities to FO frequency, signal-to-noise ratio (SNR) and observation window length. It is demonstrated that 1) the CRB of FO parameters is not affected by the presence of the transient response, 2) the CRB of the system modes is not affected by the presence of an FO in steady-state and 3) the CRB of the system modes can be drastically reduced by the presence of a FO startup transient.

\end{abstract}

\section{Introduction}
\label{sec:Intro}
The ability to estimate electromechanical modes as accurately as possible during a forced oscillation (FO) event is a critical task due to the potential for mode meters to model the FO as a system mode and throw false alarms on low damping \cite{vanfretti2012effects}. With the ever-increasing penetration of inverter-based generation and microgrids, the potential for FOs is also increasing \cite{IBRFO}, while traditional power system components continue to cause FOs such as the 2019 system-wide event in the United States Eastern Interconnection \cite{NERC}.

A highly useful tool for assessing the accuracy of an estimation algorithm is the Cram\'{e}r-Rao Bound (CRB), which defines the theoretical lower limit on the variance of an unbiased estimator. The CRB may be used as a benchmark with which to compare the variance of potential mode meters and FO parameter estimators before a particular method is selected for use. Additionally, analysis of the CRBs may yield insight into the underlying processes being estimated. Such a study was completed for mode meters under purely ambient conditions in \cite{CRLBModes}, while the authors of \cite{CRLB_FO1} and \cite{CRLB_FO2} considered the case where unknown FOs were present in a known power system. 

This paper considers the practical case where neither the FO nor the system parameters are known. After a brief review of small-signal power system modeling, the CRBs for FO parameters and mode frequency and damping are derived for two specific cases: one where only the steady-state response to the FO is present, and one that includes the FO startup transient. A simulation study is conducted that reveals several interesting observations, the most important of which is that the presence of transients associated with the onset of a FO can drastically increase the accuracy of electromechanical mode meters.

\section{System Models}
\label{sec:Background}
It is well-known that for the purposes of electromechanical mode meters and oscillation monitoring systems where preprocessed PMU data are used (e.g., lowpass filtered and detrended bus voltage angle differences), a power system under ambient conditions is well-modeled with an autoregressive moving average (ARMA) structure. When $N$ samples of $y$ are collected at sampling rate $f_s$, they are indexed from $k=0$ to $k=N-1$ with the $k^{th}$ sample of system output $y$ given as
\begin{equation}
	y[k] = \frac{C(q)}{A(q)}e[k] = \frac{1+c_1q^{-1}+\cdots+c_{n_a}q^{n_c}}{1+a_1q^{-1}+\cdots+a_{n_a}q^{n_a}}e[k]
	\label{eqn:ARMAdefn} 
\end{equation}
where $A(q)$ and $C(q)$ are the AR and MA polynomials in delay operator $q$ such that $q^{-n}y[k]=y[k-n]$, and $e$ is Gaussian White Noise (GWN) attributed to low-level random load variations. The electromechanical mode frequency and damping are computed from the system poles, i.e., the $n_a$ roots of $A(q)$. For the $i^{th}$ pole $p_i$, the modal frequency in Hz and percent damping are found as
\begin{equation}
	f_{mi} = \frac{\mathfrak{Im}(f_s \log p_i)}{2\pi}
\end{equation}
\begin{equation}
	\zeta_{mi} = -\cos\left(\phase{f_s \log p_i}\right)\times 100\%
\end{equation}

Note that in (\ref{eqn:ARMAdefn}), $e$ is the only system input being modeled. When FOs are present in $y$, the autoregressive moving average with exogenous input (ARMAX) model may be used:
\begin{equation}
	y[k] = \frac{B(q)}{A(q)}u[k] + \frac{C(q)}{A(q)}e[k]
	\label{eqn:ARMAXdefn} 
\end{equation}
where $B(q)=b_o+b_1q^{-1}+\cdots+b_{n_b}q^{n_b}$ is the X polynomial and $u$, the input FO, is in the general the sum of $p$ cosines 
\begin{equation}
	u[k] = \sum_{i=1}^p \tilde{A}_i\cos\left(\omega_i k + \tilde{\phi}_i\right)I_{\epsilon_i, \eta_i}[k]
	\label{eqn:FO}
\end{equation}
where $\tilde{A}$, $\tilde{\phi}_i$, and $\omega_i$ are amplitude, phase in radians, and frequency in radians per sample. Recall that frequency in Hz is related to radians per sample by $f_i= \omega_i f_s / (2\pi)$. Function $I$ defines the FO starting and ending samples, $\epsilon_i$ and $\eta_i$, as
\begin{equation}
	I_{\epsilon_i,\eta_i}[k] = \begin{cases}
				1, \quad \epsilon_i \leq k \leq \eta_i \\
				0, \quad \text{else} \\
			\end{cases}
	\label{eqn:I}
\end{equation} 
Note that in reality the sources of these FOs are very much a part of the power system, despite the use of the word ``exogenous'' in the ARMAX acronym. For the remainder of this paper, in order to simplify derivations and analysis, only a single FO ($p=1$) is considered. Extensions to multiple FOs are straightforward. 

The FO observed in output $y$ depends on when the FO starts and stops. This paper considers two specific cases. In the first, the FO starts long enough before the first observation of $y$, and lasts throughout the entirety of the data record. In this case, only the steady-state response of the FO is present in $y$:
\begin{equation}
	y[k] = A_1\cos(\omega_1 k + \phi_1) + \frac{C(q)}{A(q)}e[k]
	\label{eqn:FOcase1} 
\end{equation}
where amplitude and phase of the output FO are
\begin{equation}
	A_1 = \frac{|B(\omega_1)|}{|A(\omega_1)|}\tilde{A}_1
	\label{eqn:SSAmp} 
\end{equation}
\begin{equation}
	\phi_1 = \tilde{\phi}_1 + \phase{B(\omega_1)} - \phase{A(\omega_1)}
	\label{eqn:SSAng} 
\end{equation}

In the second case, the FO again lasts throughout the entirety of the data record, however it starts precisely at the first sample of $y$. Thus, $y$ includes the transient response of the FO onset:
\begin{equation}
\begin{split}
	y[k] &= \frac{B(q)}{A(q)}\tilde{A}_1\cos(\omega_1 k + \tilde{\phi}_1) + \frac{C(q)}{A(q)}e[k] \\
	     &= A_1\cos(\omega_1 k + \phi_1) + \sum_{i=1}^{n_a}  r_ip_i^k + \frac{C(q)}{A(q)}e[k] 
\end{split}	
	\label{eqn:FOcase2} 
\end{equation}
where $r_i$ are residue terms. While both (\ref{eqn:FOcase1}) and (\ref{eqn:FOcase2}) contain the same ARMA process and steady-state FO response, only (\ref{eqn:FOcase2}) contains the transient response that contains information about both the FO and the system. It is demonstrated later that this additional term can have a profound effect on the CRB of the mode meter frequency and damping estimators.

The polynomial coefficients and FO parameters of models (\ref{eqn:FOcase1}) and (\ref{eqn:FOcase2}) may be collected  in a vector termed $\theta_o$. Estimating these parameters from observations of $y$ involves minimizing a cost function of  \textit{prediction errors}, which are defined as 
\begin{equation}
	\varepsilon(k,\hat{\theta}) = y[k] - \hat{y}(k,\hat{\theta})
	\label{eqn:vareps}
\end{equation}
where $\hat{\theta}$ is an estimate of $\theta_o$ and \textit{predictor} $\hat{y}$ is the estimate of $y$ for a particular $\hat{\theta}$. The prediction errors are thus estimates of the random system input $e$. Indeed, $\varepsilon(k,\theta_o)=e[k]$.

\section{Asymptotic Cram\'{e}r-Rao Bounds}

In \cite{Ljung}, a general expression for the asymptotic Cram\'{e}r-Rao Bound of dyanimcal systems is given as
\begin{equation}
	\text{Cov}(\hat{\theta}) \geq \frac{\sigma_e^2}{N}\left[\frac{1}{N}\sum_{k=0}^{N-1}\text{E}\left\lbrace\psi(k,\theta_o)\psi^T(k,\theta_o)\right\rbrace\right]^{-1}
	\label{eqn:genCRB}
\end{equation}
where $E$ is the expected value operation, $\sigma_e^2$ is the variance of $e$, and $\psi$ is gradient
\begin{equation}
	\psi(k,\theta_o) = \frac{d}{d\theta}\hat{y}(k,\theta_o)
	\label{eqn:psi}
\end{equation}
In practice, the expected value operation in (\ref{eqn:genCRB}) must be approximated numerically by averaging over $M$ Monte Carlo simulations, each using statistically independent realization of $e$. The approximate CRB is 
\begin{equation}
	\text{Cov}(\hat{\theta}) \gtrapprox \frac{\sigma_e^2}{N}\left[\frac{1}{N}\sum_{k=0}^{N-1}\frac{1}{M}\sum_{i=1}^{M}\left\lbrace\psi_i(k,\theta_o)\psi_i^T(k,\theta_o)\right\rbrace\right]^{-1}
	\label{eqn:estCRB}
\end{equation}
where $\psi_i(k,\theta)$ is the gradient vector from the $i^{th}$ Monte Carlo trial. In practice, $M=1000$ Monte Carlo trials is more than sufficient to observe convergence in the average.

Finally, note that the electromechanical mode frequency and damping are secondary parameters. They do not appear in $\theta_o$, but are are functions of some of the elements of $\theta_o$. In order to find the CRB for these, the Taylor Series method used in \cite{CRLBModes} may be applied. 

In the following two subsections, the CRB are derived for the cases defined in (\ref{eqn:FOcase1}) and (\ref{eqn:FOcase2}).

\subsection{Case 1}
Here only the steady-state FO response is present in $y$. Rewriting expression (\ref{eqn:FOcase1}) as 
\begin{equation}
	A(q)y[k] = A(q)A_1\cos(\omega_1 k + \phi_1) + C(q)e[k]
	\label{eqn:FOcase1a} 
\end{equation}
leads to 
\begin{equation}
\begin{split}
	\hat{y}(k,\theta_o) =  - & \sum_{i=1}^{n_a}a_iy[k-i] + \sum_{i=1}^{n_c}c_i\varepsilon(k-i,\theta_o)\\
	+& \sum_{i=0}^{n_a}a_iA_1\cos(\omega_1 (k-1) + \phi_1)
\end{split}
	\label{eqn:yhatCase1}
\end{equation}
where
\begin{equation}
	\theta_o = \begin{bmatrix}
	a_1 & \cdots & a_{n_a} & c_1 & \cdots & c_{n_c} & A_1 & \phi_1 & \omega_1
	\end{bmatrix}^T
	\label{eqn:thetaCase1}
\end{equation}
The resultant gradient vector is 
\begin{equation}
\begin{split}
	\psi(k,\theta_o) = & \left[ \begin{matrix} \dfrac{\partial\hat{y}(k,\theta_o)}{\partial a_1} & \cdots & \dfrac{\partial\hat{y}(k,\theta_o)}{\partial a_{n_a}} \end{matrix} \right. \\
	& \;\;\;\;\;\;\; \begin{matrix}\dfrac{\partial\hat{y}(k,\theta_o)}{\partial c_1} & \cdots & \dfrac{\partial\hat{y}(k,\theta_o)}{\partial c_{n_c}} \end{matrix} \\
	& \left. \begin{matrix}  \;\;\;\;\;\;\;\;\; \dfrac{\partial\hat{y}(k,\theta_o)}{\partial A_1} &  \dfrac{\partial\hat{y}(k,\theta_o)}{\partial \phi_1}  & \dfrac{\partial\hat{y}(k,\theta_o)}{\partial \omega_1} \end{matrix}\right]^T
\end{split}
\end{equation}
with elements
\begin{equation}
\begin{split}
	&\frac{\partial\hat{y}(k,\theta_o)}{\partial a_i} = \frac{1}{C(q)}A_1\cos(\omega_1 (k-i) + \phi_1) \\
	&\;\;\;\;\;\;\;\;\;\;\;\;\;\;\;\;\;\;\; -\frac{1}{C(q)}y[k-i] 
\end{split}
\label{eqn:dyda11}
\end{equation}
\begin{flalign}
	&\frac{\partial\hat{y}(k,\theta_o)}{\partial c_i} = \frac{1}{C(q)}\varepsilon(k-i,\theta_o)\\[5pt]
	&\frac{\partial\hat{y}(k,\theta_o)}{\partial A_1} = \frac{A(q)}{C(q)}\cos(\omega_1 k + \phi_1)\\[5pt]
	&\frac{\partial\hat{y}(k,\theta_o)}{\partial \phi_1} = \frac{A(q)}{C(q)}(-A_1\sin(\omega_1 k + \phi_1))\\[5pt]
	&\frac{\partial\hat{y}(k,\theta_o)}{\partial \omega_1} = \frac{A(q)}{C(q)}(-A_1k\sin(\omega_1 k + \phi_1))
\label{eqn:dyda1}
\end{flalign}
where the details of the derivative calculations have been omitted due to space constraints. 

Thus, finding the CRB of $\hat{\theta}$ for a particular ARMA model and FO involves the following steps. First, $M$ independent $N$-sample sequences of $e$ are generated and used to create $M$ realizations of $y$. These, along with the sinusoids in (\ref{eqn:dyda11}) -  (\ref{eqn:dyda1}), are filtered through $1/C(q)$ or $A(q)/C(q)$ to obtain the $M$ realizations of $\psi$ that are used by (\ref{eqn:estCRB}) to obtain the CRB. A subtle but very important detail here is that since $y$ only contains steady-state FO responses, care must be taken to ensure that only the steady-state responses to the filtering operations in (\ref{eqn:dyda11}) -  (\ref{eqn:dyda1}) are included in the gradient vectors.

The CRB of $A_1$, $\phi_1$ and $\omega_1$ are found as the final three diagonal elements of the CRB of $\hat{\theta}$, and the CRB of the mode frequency and damping are obtained by applying the the CRB of the AR polynomial coefficients (the $n_a\times n_a$ upper left block of the CRB of $\hat{\theta}$) to the methods found in \cite{CRLBModes}.

\subsection{Case 2}
Rewriting (\ref{eqn:FOcase2}),
\begin{equation}
\begin{split}
	\hat{y}(k,\theta_o) =  - & \sum_{i=1}^{n_a}a_iy[k-i] + \sum_{i=1}^{n_c}c_i\varepsilon(k-i,\theta_o)\\
	+& \sum_{i=0}^{n_b}b_i\tilde{A}_1\cos(\omega_1 (k-1) + \tilde{\phi}_1)
\end{split}
	\label{eqn:yhatCase2}
\end{equation}
with
\begin{equation}
\begin{split}
	\theta_o = & \left[ \begin{matrix} a_1 & \cdots & a_{n_a} & b_0 & \cdots & b_{n_b} \end{matrix} \right. \\
	&\;\;\;\;\;\;\;\;\;\;\;\;\;\;\; \left. \begin{matrix} c_1 & \cdots & c_{n_c} & \tilde{A}_1 & \tilde{\phi}_1 & \omega_1 	\end{matrix} \right]^T
\end{split}
	\label{eqn:thetaCase2}
\end{equation}
and 
\begin{equation}
\begin{split}
	\psi(k,\theta_o) = & \left[ \begin{matrix} \dfrac{\partial\hat{y}(k,\theta_o)}{\partial a_1} & \cdots & \dfrac{\partial\hat{y}(k,\theta_o)}{\partial a_{n_a}} \end{matrix} \right. \\
	& \;\;\;\;\; \begin{matrix}\dfrac{\partial\hat{y}(k,\theta_o)}{\partial b_0} & \cdots & \dfrac{\partial\hat{y}(k,\theta_o)}{\partial b_{n_b}} \end{matrix} \\
	& \;\;\;\;\;\;\; \begin{matrix}\dfrac{\partial\hat{y}(k,\theta_o)}{\partial c_1} & \cdots & \dfrac{\partial\hat{y}(k,\theta_o)}{\partial c_{n_c}} \end{matrix} \\
	& \left. \begin{matrix}  \;\;\;\;\;\;\;\;\; \dfrac{\partial\hat{y}(k,\theta_o)}{\partial \tilde{A}_1} &  \dfrac{\partial\hat{y}(k,\theta_o)}{\partial \tilde{\phi}_1}  & \dfrac{\partial\hat{y}(k,\theta_o)}{\partial \omega_1} \end{matrix}\right]^T
\end{split}
\end{equation}
with elements
\begin{equation}
\begin{split}
	&\frac{\partial\hat{y}(k,\theta_o)}{\partial a_i} = \frac{1}{C(q)}\left(-y[k-i]  \right) \;\;\;\;\;\;\;\;\;\;\;\;\;\;\;\;
\end{split}
\label{eqn:dyda21}
\end{equation}
\begin{flalign}
	&\frac{\partial\hat{y}(k,\theta_o)}{\partial b_i} = \frac{1}{C(q)} \tilde{A}_1\cos(\omega_1 (k-i) + \tilde{\phi}_1)\\[5pt]
	&\frac{\partial\hat{y}(k,\theta_o)}{\partial c_i} = \frac{1}{C(q)}\varepsilon(k-i,\theta_o)\\[5pt]
	&\frac{\partial\hat{y}(k,\theta_o)}{\partial \tilde{A}_1} = \frac{B(q)}{C(q)}\cos(\omega_1 k + \tilde{\phi}_1)\\[5pt]
	&\frac{\partial\hat{y}(k,\theta_o)}{\partial \tilde{\phi}_1} = \frac{B(q)}{C(q)}(-\tilde{A}_1\sin(\omega_1 k + \tilde{\phi}_1))\\[5pt]
	&\frac{\partial\hat{y}(k,\theta_o)}{\partial \omega_1} = \frac{B(q)}{C(q)}(-\tilde{A}_1k\sin(\omega_1 k + \tilde{\phi}_1))
\label{eqn:dyda2}
\end{flalign}

Similar to Case 1, $M$ independent realizations of $e$ are created and applied to (\ref{eqn:FOcase2}) to generate $M$ independent realizations of $y$, both of which are filtered by $1/C(q)$ or $B(q)/C(q)$ along with the sinusoids of (\ref{eqn:dyda21}) -  (\ref{eqn:dyda2}) to create $M$ realizations of $\psi$ that are used by (\ref{eqn:estCRB}) to obtain the CRB. Here, the subtle but important note is that since $y$ does contain the start-up transient responses of the FO, so to should the gradient elements. 

Note that in this case, while $\theta_o$ contains the FO frequency common to both the input and output FO, it only contains the amplitude and phase of the \textit{input} FO. Thus, while the CRB of $\omega_1$ is found as the last diagonal element of the CRB of $\hat{\theta}$, the CRB of $A_1$ and $\phi_1$ must be obtained by applying Taylor Series linearizations to the CRB of $\hat{\theta}$. 

Referring back to (\ref{eqn:SSAmp}) and (\ref{eqn:SSAng}), define $X$ as the phasor representation of the output FO
\begin{equation}
	X = A_1e^{j\phi_1} = \frac{B(\omega_1)}{A(\omega_1)}\tilde{A}_1e^{j\tilde{\phi}_1} = \alpha + j\beta
\end{equation}
where $\alpha$ and $\beta$ are the real and imaginary parts of $X$. First, the CRB of $\begin{bmatrix} \alpha & \beta\end{bmatrix}^T$ is found from the CRB of $\hat{\theta}$ as 
\begin{equation}
	\text{Cov}\begin{bmatrix}	\alpha \\ \beta	\end{bmatrix} = J_{\alpha\beta}\text{Cov}(\hat{\theta)} J_{\alpha\beta}^T
\end{equation}
with the Jacobian obtained from the real and imaginary parts of complex Jacobian $J_X$
\begin{equation}
	J_{\alpha\beta} = \begin{bmatrix}
	\mathfrak{Re}(J_X)\\ \mathfrak{Im}(J_X)
	\end{bmatrix}
\end{equation}
where 
\begin{equation}
\begin{split}
	J_X &= \left[ \begin{matrix} \dfrac{\partial X}{\partial a_1} & \cdots & \dfrac{\partial X}{\partial a_{n_a}} & \dfrac{\partial X}{\partial b_0} & \cdots & \dfrac{\partial X}{\partial b_{n_b}}   \end{matrix} \right. \\
	& \;\;\;\;\;\;\;\;\;\;\; \left. \begin{matrix}  \dfrac{\partial X}{\partial c_1} & \cdots & \dfrac{\partial X}{\partial c_{n_c}} & \dfrac{\partial X}{\partial \tilde{A}_1} &  \dfrac{\partial X}{\partial \tilde{\phi}_1}  & \dfrac{\partial X}{\partial \omega_1} \end{matrix}\right]
\end{split}
\label{eqn:Jx}
\end{equation}
with elements
\begin{flalign}
	&\frac{\partial X}{\partial a_i} = \frac{-e^{-ji\omega_1}B(\omega_1)}{A^2(\omega_1)}\tilde{A}_1e^{j\tilde{\phi}_1} \\[5pt]
	&\frac{\partial X}{\partial b_i} = \frac{e^{-ji\omega_1}}{A(\omega_1)}\tilde{A}_1e^{j\tilde{\phi}_1} \\[5pt]
	&\frac{\partial X}{\partial c_i} = 0\\[5pt]
	&\frac{\partial X}{\partial \tilde{A}_1} = \frac{B(\omega_1)}{A(\omega_1)}e^{j\tilde{\phi}_1} \\[5pt]
	&\frac{\partial X}{\partial \tilde{\phi}_1} = \frac{B(\omega_1)}{A(\omega_1)}j\tilde{A}_1e^{j\tilde{\phi}_1}\\[5pt]\nonumber
	&\frac{\partial X}{\partial \omega_1} = \frac{B(\omega_1)\sum\limits_{i=0}^{n_a}ia_ie^{-ji\omega_1} - A(\omega_1)\sum\limits_{i=0}^{n_b}ib_ie^{-ji\omega_1}}{A^2(\omega_1)} \\[5pt]	
	&\;\;\;\;\;\;\;\;\;\;\; \times\tilde{A}_1e^{j\tilde{\phi}_1}
\label{eqn:JacElem}
\end{flalign}
The CRB of $\begin{bmatrix} A_1 & \phi_1\end{bmatrix}^T$  is then found from the CRB of $\begin{bmatrix} \alpha & \beta\end{bmatrix}^T$ as 
\begin{equation}
	\text{Cov}\begin{bmatrix}	A_1 \\ \phi_1	\end{bmatrix} = J_{A\phi}\text{Cov}\begin{bmatrix}	\alpha \\ \beta	\end{bmatrix} J_{A\phi}^T
	\label{eqn:CRBAphi}
\end{equation}
with Jacobian
\begin{equation}
	J_{A\phi} = 	
	\begin{bmatrix}
		\dfrac{\partial A_1}{\partial\alpha}     &   \dfrac{\partial A_1}{\partial\beta}    \\
		\dfrac{\partial \phi_1}{\partial\alpha}  &   \dfrac{\partial \phi_1}{\partial\beta}
	\end{bmatrix}
	=
	\frac{1}{A_1}\begin{bmatrix}
		A_1\cos(\phi_1)                             &   A_1\sin(\phi_1) \\
		-\sin(\phi_1)              &   \cos(\phi_1)
	\end{bmatrix}	
	\label{eqn:Japhi}
\end{equation}
Finally, the CRBs of $A_1$ and $\phi_1$ are the diagonal elements of (\ref{eqn:CRBAphi}), and as with Case 1, the CRB of the mode frequency and damping are obtained from applying the Taylor Series methods of \cite{CRLBModes} to the $n_a\times n_a$ upper left block of the CRB of $\hat{\theta}$).

\section{Numerical Study}
\begin{figure}[!t]
	\centering
	\includegraphics[width=2.7in]{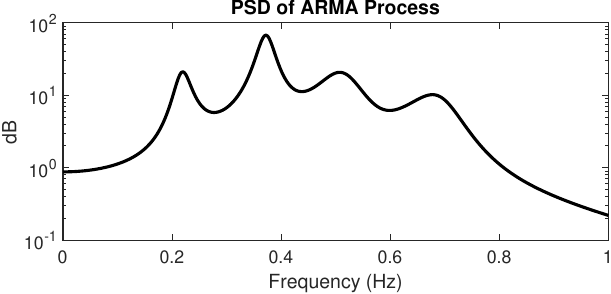}
	\caption{Power spectral density of the ARMA system used in the simulations. }
	\label{fig:Fig1}	
	\vspace{-0.1in}
\end{figure}

To illustrate how the parameters of a FO may affect the estimation accuracy of FO estimation tools and electromechancial mode meters, a numerical study was conducted using a low-order ARMAX approximation of the minniWECC model of the Western Electricity Coordinating Council (WECC) power system. Details on the full minniWECC model are found in \cite{miniwecc}. As was initially demonstrated in \cite{CRLBModes}, when a single output of the minniWECC model is preprocessed by detrending, lowpass filtering and downsampling from 120 to 3 samples per second, only a few of the system dynamics remain, which are very well-modeled as a low-order ARMAX system. Here, simulations were conducted with AR and MA orders of 10, and an X order of 1. The power spectral density (PSD) of the ARMA portion of the system is shown in Fig. \ref{fig:Fig1}, where the local maxima correspond to system modes. Here the main North-South interea mode at 0.372 Hz and 4.67 \% damping was considered for estimation.

Four simulation scenarios were considered. For all scenarios, the output FO amplitude and phase were $A_1=1$ and $\phi_1=0.8$ rad ($45.8^{\circ}$). In the first, the CRBs were calculated over a range of FO frequencies with a record length of 5400 samples (30 minutes), and noise variance $\sigma_e^2$ that maintained a constant local signal-to-noise ratio (SNR) of 40 dB. Local SNR is defined with the output FO as the signal and the PSD of the ARMA process at the FO frequency as the noise. The second scenario is identical to the first, with the exception that a constant SNR of 9.5 dB is used, where SNR takes its classic definition of the noise power defined as the average PSD of the ARMA process across the entire frequency range. In the third scenario, FO frequency is held at 0.353 Hz with a 30-minute record length while the CRBs were found for a variety of SNR. Finally, in the fourth scenario, FO frequency and SNR are held at 0.353 Hz and 9.5 dB while the CRBs were found for a range of record lengths. In each of these scenarios, both of the previously discussed FO cases were studied.

\begin{figure}[!t]
	\centering
	\includegraphics[width=2.9in]{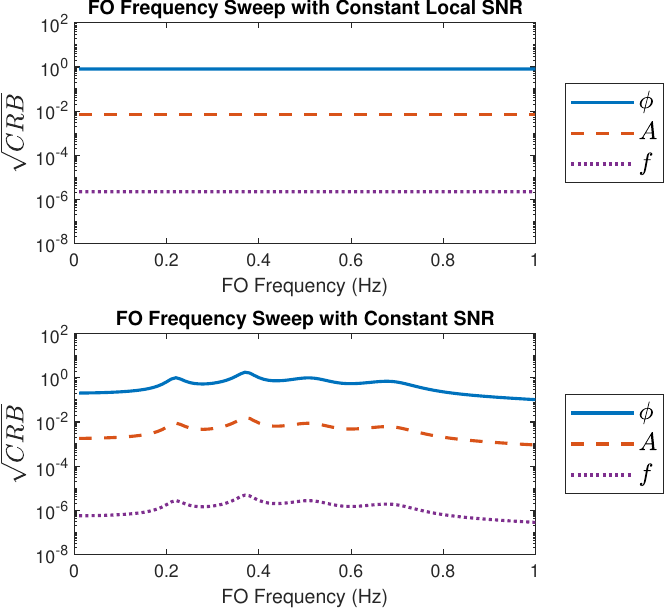}
	\caption{Square root CRB for FO parameters as a function of FO frequency.}
	\label{fig:Fig2}	
	\vspace{-0.1in}
\end{figure}

\begin{figure}[!t]
	\centering
	\includegraphics[width=2.9in]{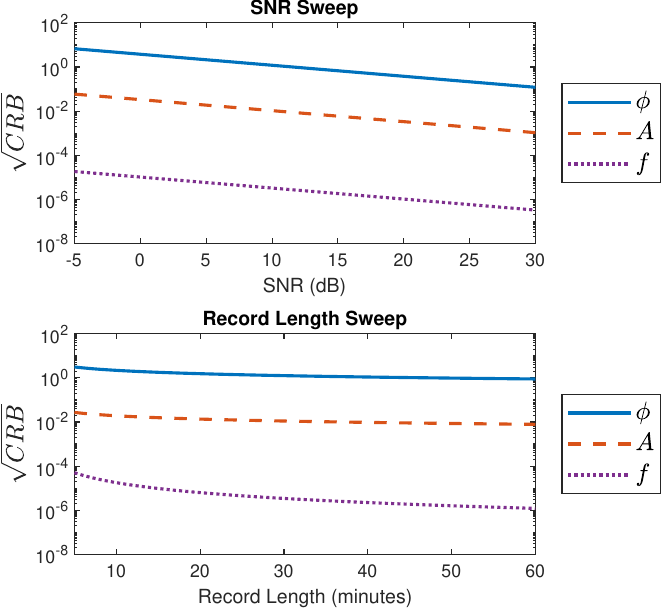}
	\caption{Square root CRB for FO parameters as a function of global SNR (top) and record length (bottom). }
	\label{fig:Fig3}	
	\vspace{-0.1in}
\end{figure}

Results for the FO parameters are shown in Figs. \ref{fig:Fig2} and \ref{fig:Fig3}. Note that these plots  were virtually identical for each of the two FO cases, indicating that \textbf{the presence of the FO transient transient response has no effect on the ability to estimate its steady-state parameters.} The frequency sweeps of Fig. \ref{fig:Fig2} illustrate that the CRBs are independent of FO frequency if the local SNR is held constant, while they follow the pattern of ARMA PSD shown in Fig. \ref{fig:Fig1} if the overall SNR is held constant. The results of Fig. \ref{fig:Fig3} that the CRBs of all three FO parameters decreases as SNR and record length increase, albeit at different rates.

\begin{figure}[!t]
	\centering
	\includegraphics[width=3.3in]{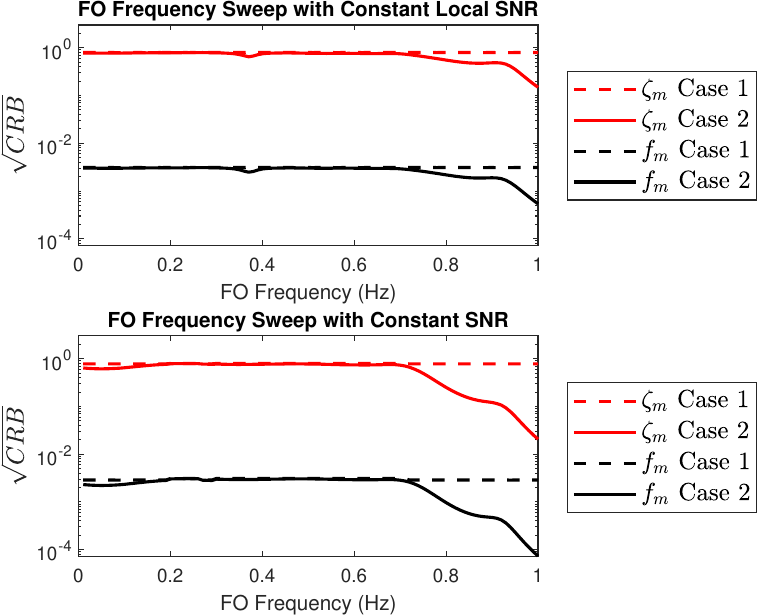}
	\caption{Square root CRB for the electromechanical mode frequency and damping as a function of FO frequency.  }
	\label{fig:Fig4}	
	\vspace{-0.1in}
\end{figure}

\begin{figure}[!t]
	\centering
	\includegraphics[width=3.3in]{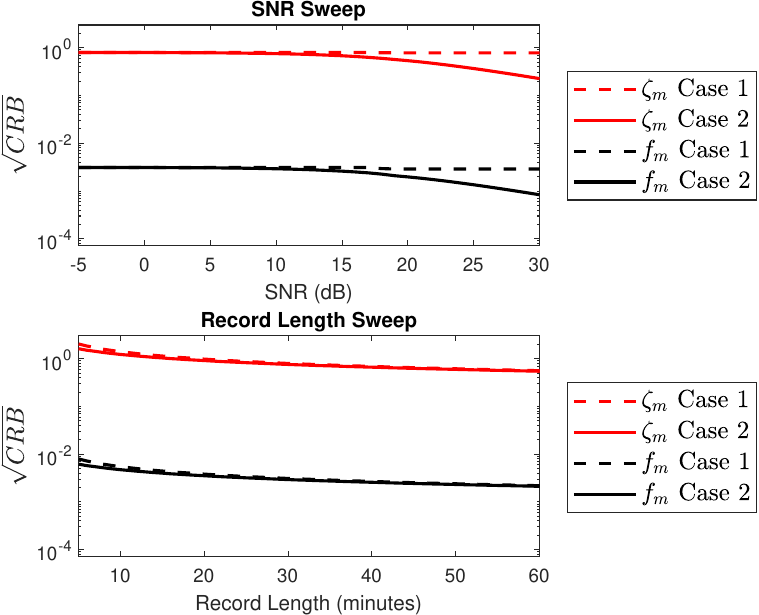}
	\caption{Square root CRB for electromechanical mode frequency and damping  as a function of global SNR (top) and record length (bottom). }
	\label{fig:Fig5}	
	\vspace{-0.15in}
\end{figure}

Results for the CRBs of the mode frequency and damping are shown in \ref{fig:Fig4} and \ref{fig:Fig5}. The first thing to note is that in case 1 the CRBs are constant and identical across the first three scenarios, and show a decay in $1/N$ in the final scenario. Indeed these results are virtually identical to the CRBs in the purely ambient case when no FO is present. This implies that \textbf{when only the FO steady-state is present in the measured output, the FO has negligible effect on the CRB of the mode meter.} 

Just as interesting are the case 2 results, which first illustrate that if the FO frequency is either very low or quite near the mode frequency, a slight decrease in CRB can be enjoyed, while if the FO frequency is higher than the electromechanical mode range, a significant decrease in CRB is seen. Similarly, a substantial drop in CRB is seen for high SNR. This indicates that \textbf{when the start-up transient of the FO is present in the measured output, increases in estimation accuracy can be achieved.}

\section{Conclusions}
\label{sec:Conclusions}
Through the derivation and analysis of the CRB on the variance of FO parameters and electromechanical mode frequency and damping, several interesting observations are noted. Arguably the most important is that whenever possible, one should include system transients due to the onset of FOs in the data to be analyzed by FO and mode estimation algorithms. Ongoing and future work in this area include determining the effects of including FO ending transients in analysis window, along with using these CRBs as benchmarks to test the variance for FO estimation and meter algorithms, especially in cases where the FO parameters and possibly even the system modes are slowly changing.

\bibliographystyle{IEEEtran}
\bibliography{IEEEabrv,PES_GM_2024}

\begin{thebibliography}{1}
\providecommand{\url}[1]{#1}
\csname url@samestyle\endcsname
\providecommand{\newblock}{\relax}
\providecommand{\bibinfo}[2]{#2}
\providecommand{\BIBentrySTDinterwordspacing}{\spaceskip=0pt\relax}
\providecommand{\BIBentryALTinterwordstretchfactor}{4}
\providecommand{\BIBentryALTinterwordspacing}{\spaceskip=\fontdimen2\font plus
\BIBentryALTinterwordstretchfactor\fontdimen3\font minus
  \fontdimen4\font\relax}
\providecommand{\BIBforeignlanguage}[2]{{%
\expandafter\ifx\csname l@#1\endcsname\relax
\typeout{** WARNING: IEEEtran.bst: No hyphenation pattern has been}%
\typeout{** loaded for the language `#1'. Using the pattern for}%
\typeout{** the default language instead.}%
\else
\language=\csname l@#1\endcsname
\fi
#2}}
\providecommand{\BIBdecl}{\relax}
\BIBdecl

\bibitem{vanfretti2012effects}
L.~Vanfretti, S.~Bengtsson, V.~S. Peri{\'c}, and J.~O. Gjerde, ``Effects of
  forced oscillations in power system damping estimation,'' in \emph{Applied
  Measurements for Power Systems (AMPS), 2012 IEEE International Workshop
  on}.\hskip 1em plus 0.5em minus 0.4em\relax IEEE, 2012, pp. 1--6.

\bibitem{IBRFO}
C.~Wang, C.~Mishra, K.~D. Jones, R.~M. Gardner, and L.~Vanfretti, ``Identifying
  oscillations injected by inverter-based solar energy sources,'' in \emph{2022
  IEEE Power \& Energy Society General Meeting (PESGM)}, 2022, pp. 1--5.

\bibitem{NERC}
A.~Alam~et al., ``Eastern interconnection oscillation disturbance januarry 11,
  2019 forced oscillation event,'' North American Electric Reliability
  Corporation (NERC), Tech. Rep., December 2019.

\bibitem{CRLBModes}
L.~Dosiek, ``On the cramer–rao bound of power system electromechanical mode
  meters,'' \emph{IEEE Transactions on Power Systems}, vol.~34, no.~6, pp.
  4674--4683, 2019.

\bibitem{CRLB_FO1}
Z.~Xu and J.~W. Pierre, ``Initial results for cramer-rao lower bound for forced
  oscillations in power systems,'' in \emph{2021 North American Power Symposium
  (NAPS)}, 2021, pp. 1--5.

\bibitem{CRLB_FO2}
Z.~Xu, J.~W. Pierre, R.~Elliott, D.~Schoenwald, F.~Wilches-Bernal, and
  B.~Pierre, ``Cramer-rao lower bound for forced oscillations under
  multi-channel power systems measurements,'' in \emph{2022 17th International
  Conference on Probabilistic Methods Applied to Power Systems (PMAPS)}, 2022,
  pp. 1--6.

\bibitem{Ljung}
L.~Ljung, \emph{System Identification: Theory for the User}, 2nd~ed.\hskip 1em
  plus 0.5em minus 0.4em\relax Upper Saddle River, NJ: Prentice Hall, 1999.

\bibitem{miniwecc}
D.~Trudnowski, D.~Kosterev, and J.~Undrill, ``{PDCI} damping control analysis
  for the western north american power system,'' in \emph{2013 IEEE Power
  Energy Society General Meeting}, July 2013, pp. 1--5.

\end{thebibliography}

\end{document}